\documentclass[prd,aps,amsmath,amssymb,nofootinbib,twocolumn,preprintnumbers]
{revtex4}

\usepackage{graphicx}
\usepackage{dcolumn}
\usepackage{bm}
\usepackage{amsmath}
\usepackage{amsthm}
\usepackage{amsfonts}
\usepackage{euscript,bbm}
\usepackage{ifthen}
\usepackage{psfrag}
\usepackage{slashed}
\usepackage{hyperref}
\usepackage{color}

\def\ls{\mathrel{\lower4pt\vbox{\lineskip=0pt\baselineskip=0pt
           \hbox{$<$}\hbox{$\sim$}}}}
\def\gs{\mathrel{\lower4pt\vbox{\lineskip=0pt\baselineskip=0pt
           \hbox{$>$}\hbox{$\sim$}}}}
\def\drawbox#1#2{\hrule height#2pt
\hbox{\vrule width#2pt height#1pt \kern#1pt
              \vrule width#2pt}
              \hrule height#2pt}

\def\Asym#1#2{\vcenter{\vbox{\drawbox{#1}{#2}
              \kern-#2pt       % line up boxes
              \drawbox{#1}{#2}}}}

\def\nn{\nonumber}

\newcommand{\be}{\begin{equation}}
\newcommand{\ee}{\end{equation}}
\newcommand{\bea}{\begin{eqnarray}}
\newcommand{\eea}{\end{eqnarray}}

\newcommand{\gsim}{\lower.7ex\hbox{$\;\stackrel{\textstyle>}{\sim}\;$}}
\newcommand{\lsim}{\lower.7ex\hbox{$\;\stackrel{\textstyle<}{\sim}\;$}}

\newcommand{\sci}[2]{#1$\times$10$^{\text{#2}}$}

\newcommand{\ben}{\begin{enumerate}}
\newcommand{\een}{\end{enumerate}}
\newcommand{\bei}{\begin{itemize}}
\newcommand{\eei}{\end{itemize}}

\begin{document}

MI-TH-1509

\title{The Leptoquark Implication from the CMS and IceCube Experiments}

\author{Bhaskar Dutta$^{1}$}
\author{Yu Gao$^{1}$}
\author{Tianjun Li$^{2,3}$}
\author{Carsten Rott$^{4}$}
\author{Louis E. Strigari$^{1}$}
\affiliation{
$^{1}$~Department of Physics and Astronomy, Mitchell Institute for Fundamental Physics and Astronomy, Texas A\&M University, College Station, TX 77843-4242, USA\\
$^{2}$State Key laboratory of Theoretical Physics and Kavli Institute for Theoretical Physics China (KITPC), Institute of Theoretical Physics, Chinese Academy of Sciences, Beijing 100190, China\\
$^{3}$School of Physical Electronics, University of Electronic Science and Technology of China, Chengdu 610054, China\\
$^{4}$Department of Physics, Sungkyunkwan University, Suwon 440-746, Korea
}

\begin{abstract}
The recent excess in the CMS measurements of $eejj$ and $e\nu jj$ channels and the emergence of PeV cosmic neutrino events 
at the IceCube experiment share an intriguing implication for a leptoquark with a 600-650 GeV mass. We investigate the CMS 
constraints on the flavor structure of a scenario with the minimal leptoquark Yukawa couplings and correlate such a scenario 
to the resonant enhancement in the very high energy shower event rates at the IceCube. We find for a single leptoquark, 
the CMS signals require large couplings to the third generation leptons. This leads to an enhancement in the $\nu_\tau$-nucleon 
scattering cross-section and subsequently more $\nu_\tau$ events at PeV energies. However, a visible enhancement above 
the Standard Model scattering would require a leptoquark Yukawa coupling larger than one that can be easily tested at the upcoming LHC runs.
\end{abstract}

\maketitle

\section{Introduction}

CMS has measured~\cite{bib:CMSleptoquark} a 2.4$\sigma$ excess in $eejj$ and $e\nu jj$ channels, compared to the earlier null results from 
the CMS~\cite{bib:CMSnul} and ATLAS~\cite{bib:ATLASnul} Collaborations. The CMS excess has indicated a best fit for
a leptoquark with 650 GeV mass~\cite{bib:CMSleptoquark}. This interesting possibility has been investigated 
in recent works~\cite{bib:leptoquark_refs}. Through its recent observation of high energy neutrinos~\cite{Aartsen:2013bka,Aartsen:2015ivb}, 
the IceCube has potential to test leptoquark models. Interestingly, at the PeV energy range, the neutrino flux (up to uncertainty 
due to the limited statistics) may be fit within the picture of a leptoquark resonance~\cite{Barger:2013pla}. Additional 
explanations to the IceCube neutrino flux include a new or modified cosmic source flux~\cite{bib:newSource}, new neutrino flavor(s)~\cite{bib:newFlavor}, 
colored-neutrino resonances~\cite{Akay:2014tga}, a Glashow resonance at the next-to-leading order~\cite{Alikhanov:2015kla}, Standard Model 
explanations~\cite{bib:SMexp}, as well as other possibilities~\cite{bib:otherTh}.  

In the Grand Unified Theories (GUTs)~\cite{bib:GUTS} such as $SU(5)$, $SO(10)$, and $E_6$, etc,
the lepton ($L$) and baryon ($B$) number symmetries are violated explicitly,
although the $B-L$ symmetry can still be preserved. Thus, leptoquarks  generically exist which mediate transitions between  quarks and leptons. 
In addition, these leptoquarks will not contribute to gauge anomalies since they are scalars. So they provide  simple extensions
to the Standard Model (SM). 

This paper assumes a single scalar leptoquark that explains both signals from the IceCube and the CMS, and shows that these current experimental 
hints demand a specific flavor structure in the leptoquark hypothesis, which can make predictions  for the future PeV events at the IceCube and will be 
probed efficiently by the upcoming LHC data.

Section~\ref{sect:model} gives the minimal flavor assignment for the leptoquark couplings and possible constraints. Section~\ref{sect:cms} 
and~\ref{sect:icecube} present the leptoquark's signal rates at the CMS and the IceCube. We then conclude in Section~\ref{sect:conclusion}.

\section{A minimal leptoquark scenario}
\label{sect:model}

We follow the notations of Ref.~\cite{Buchmuller:1986zs} and focus on a single scalar `S1' type leptoquark $S$
whose $SU(3)_C\times SU(2)_L \times U(1)_Y$ quantum number is $(\mathbf{\overline 3}, \mathbf{1}, \mathbf{1/3})$.
Its Yukawa couplings to the SM quarks and leptons are
\begin{eqnarray}
{\cal L} &\supset& y^L_{ij}\bar{q}^c_{i,L}i{\bf \tau}_2 l_{j,L} S +y^R_{ij} \bar{u}_{i,R}^c l_{j,R} S 
+ y^{QQ}_{ij}\bar{q}^c_{i,L}i{\bf \tau}_2 q_{j,L} S^{\dagger} 
\nonumber \\ &&
+ y^{UD}_{ij} \bar{u}_{i,R}^c d_{j,R} S^{\dagger}
+ c.c.~,~\,
\end{eqnarray}
where the subscript $i,j$ denote the flavor generations of quarks and leptons. The matrix $i{\bf \tau}_2$ swaps the two isospin elements in $l_L$. To solve the proton decay problem, consider the $Z_3$ symmetry as follows
\begin{eqnarray}
&& q_{i, L} \rightarrow \omega^2 q_i~,~~u_{i, R} \rightarrow \omega^2 u_{i, R} ~,~~
 d_{i, R} \rightarrow \omega^2 d_{i, R}
\nonumber \\ &&
l_i  \rightarrow \omega l_i~,~~e_{i, R} \rightarrow \omega e_{i, R} ~,~\,
\end{eqnarray}
where $\omega^3=1$.
Thus, we forbid the diquark couplings $ y^{QQ}_{ij}$ and $y^{UD}_{ij}$, and then escape the constraints from 
the proton decay experiments.

While a leptoquark can generally couple to all three generations of quark and lepton flavors, a minimal coupling scenario can be favored:

(i) The IceCube neutrino events occurs via neutrino scattering off the nucleons in ice, which demands left-handed coupling to the valence (first generation) quarks.

(ii) Due to the high energy nature of  incoming cosmic ray neutrinos, an $s$-channel leptoquark resonance cross-section is suppressed due to the contribution to the leptoquark decay width from any couplings other than that in (i), i.e., all right-handed couplings $y^R$ and any left-handed $y^L$ to the second or third generation quarks.

(iii) The lack of muon tracks among the highest energy events disfavors its couplings to $\mu$ and $\nu_\mu$.

(iv) The leptoquark Yukawa couplings to heavy quarks introduce constraints  arising from $l\rightarrow l^{\prime}\gamma$ decays.

Keeping the conditions (i)- (iv) in mind, we require our minimal flavor structure to consist of nonzero $y^L_{13}$ and $y^R_{21}$. The leptoquark interaction Lagrangian is 
\bea 
{\cal L}_{LQ} &=& y^L_{13}\bar{q}^c_{1,L}i{\bf \tau}_2 l_{3,L} S +y^R_{21} \bar{c}_{R}^c e_{R} S \nonumber \\
&+& y^R_{23} \bar{c}_{R}^c \tau_{R} S + {\cal O}({\epsilon}) + c.c.~,~\,
\label{eq:couplings}
\eea
where $q_1=(u,d),l_1=(\nu_e,e)$ and $l_3=(\nu_\tau,\tau)$. $y_{13}^L$ is responsible for the $\nu_\tau$-nucleon collision while $y_{21}^R,y_{23}^R$ must be included to balance the CMS signal rates in the $eejj$ and $e\nu jj$ channels.  $y^R$ does not mediate neutrino-nucleon scattering. %yet a large $y^R$ can generally suppress the IceCube rate.
We choose the right-handed charm quark in stead of up quark so that
the right-handed couplings do not cause a large enhancement to 
the rare $\pi^+ \to e^+ \nu$ decay~\cite{Davidson:1993qk} and a large radiative 
correction to the up quark mass, when the left-handed coupling to 
the 1st generation quark doublet is also present. The couplings in Eq.~\ref{eq:couplings} can lead to $D^+\rightarrow e^+\nu$ decay, yet given the uncertianty in the $D^+$ meson form factor~\cite{Davidson:1993qk}, it can tolerate relatively large couplings $y\sim 1$. ${\cal O}(\epsilon)$ denotes for other terms with couplings that are negligible in the quadrature sum $\sum_{ij} (2|y^L|^2_{ij}+|y^R|^2_{ij})$. To generate an $eejj$ signal, in principle there can also be $y_{11}^L$. However $y_{11}^L$ is also not as effective as $y^R_{21}$ to explain the CMS signal ratio which will be shown shortly.

\section{The Collider signals}
\label{sect:cms}

The CMS experiment has observed~\cite{bib:CMSleptoquark} 36 $eejj$ and 18 $e\nu jj$ events at 19.6$fb^{-1}$ whereas $20.49\pm 2.14(stat)\pm 2.45(sys)$ and $7.54\pm 1.20(stat) \pm 1.07(sys)$ events are expected from the standard model (SM) background for these final states respectively.  A best-fit with 650 GeV leptoquark would require about 20\% decay branching ratio (BR) in the $S\rightarrow ej$ channel, subject to possible variation in the leptoquark pair production cross-section at the next-to-leading order~\cite{bib:CMSleptoquark}\footnote{As a note, in case of a very large leptoquark couplings to the 1st generation quarks, t-channel lepton diagrams should also be included beside QCD processes.} (NLO) and the experimental event selection cut efficiencies. For slightly lighter leptoquarks, the BR$_{ej}$ can be lower due to a greater production rate.

Converting the CMS event rates into signal cross-sections, we get
\bea
\sigma^{LQ}_{eejj}&=& 0.79 \pm 0.33(stat) \pm 0.13(sys) \text{\ \ fb} ~,~\, \nn \\
\sigma^{LQ}_{e\nu jj}&=&0.53 \pm 0.22 (stat) \pm 0.055 (sys) \text{\ fb} ~,~\, \\
r_{e\nu}&\equiv&\frac{\sigma_{eejj}}{\sigma_{e\nu jj}}=\frac{1}{2}\frac{\text{BR}^2_{ej}}{\text{BR}_{\nu j}\text{BR}_{ej}}\frac{A_{eejj}}{A_{e\nu jj}}=1.5\pm 0.9 ~,~\, \nn
\eea
where we have added the subdominant systematics in quadrature for $r_{e\nu}$. $A$ denotes the cut efficiency in each channel. We find both $A_{eejj}$ and $A_{e\nu jj}\sim 60\%$ using {\it MadGraph v5} and electron/jet reconstruction by {\it Pythia+PGS4} for the selection cuts listed in Ref.~\cite{bib:CMSleptoquark}. $\tau$ leptons only have a few percent chance to fake an electron event and can be neglected. Ignoring the quark and lepton masses, the leptoquark decay width is
\be  
\Gamma = \frac{M_{LQ}}{16\pi}\sum_{i,j}2(y^L_{ij})^2+(y^R_{ij})^2~,~\,
\ee
where the factor of 2 arises from the left-handed coupling giving rise to both $S\rightarrow lq$ and $S\rightarrow \nu_l q'$ decays. With a minimal set of $\{y_{13}^L, y^{L}_{11},y^{R}_{21} \}$ couplings, we have
\bea 
r_{e\nu} &=& \frac{1}{2}\times \frac{(y^L_{11})^2+(y^R_{21})^2}{(y^L_{11})^2+(y^L_{13})^2},\\
\text{BR}_{ej}&=&\frac{(y^L_{11})^2+(y^R_{21})^2}{2(y^L_{11})^2+2(y^L_{13})^2+(y^R_{21})^2+(y^R_{23})^2}.
\label{eq:ratioBR}
\eea
Both $y_{11}^L$ and  $y_{21}^R$ terms contributes to BR$_{ej}$. Since the CMS has observed less signal $e\nu jj$ than $e\nu jj$, 
from Eq.~(\ref{eq:ratioBR}) it is clear that the $y_{21}^R$ term is more favored compared to the $y_{11}^L$ term which makes it  difficult to keep BR$_{\nu j}$ down. We, therefore, drop $y_{11}^L$ from now on. Although the central value prefers a larger $y^R_{21}$, due to the still significant statistical uncertainty, $y^R_{21} \sim y^L_{13}$ is still allowed, and two  
benchmark points are given in Table~\ref{tab:benchmarks}.

\begin{table}[h]
\begin{tabular}{c|cc|ccc}
\hline
$M_{LQ}$& $|y^R_{21}/y^L_{13}|^2$  & $|y^R_{23}/y^L_{13}|^2$ & BR$_{ej}$& BR$_{\nu j}$ &BR$_{\tau j}$ \\
\hline
600 & 1 & 2 & 20\%& 20\%& 60\% \\
650 & 0.57 & 0.29 & 20\%& 35\%& 45\% \\
\hline
\end{tabular}
\caption{The benchmark points for the leptoquark Yukawa coupling assignment schemes that are consistent with the CMS measurements.}
\label{tab:benchmarks}
\end{table}

The scalar leptoquark pair production is QCD dominant, and its next to leading order (NLO) cross-section is given in Ref.~\cite{Kramer:2004df}. At the benchmark points, a 600(650) GeV leptoquark has a cross-section of $\sigma_0=20(11)$ fb. With a 60\% selection cut efficiency, it leads to the after-cut signal $\sigma_{eejj}=0.5(0.3)$ fb and $\sigma_{e\nu jj}=0.96(0.9)$ fb, within 2$\sigma$ uncertainty of the CMS observed event rates.

It is worthwhile to notice that a sizable $y_{23}^R$ is needed to prevent $y_{13}^L$ from dominating all of the non-$ej$ decay branching ratio, in which case a large signal $\sigma_{e\nu jj}$ may easily be ruled out by the current CMS measurement.

\section{Icecube neutrino events}
\label{sect:icecube}

\begin{figure}[h]
\includegraphics[scale=0.7]{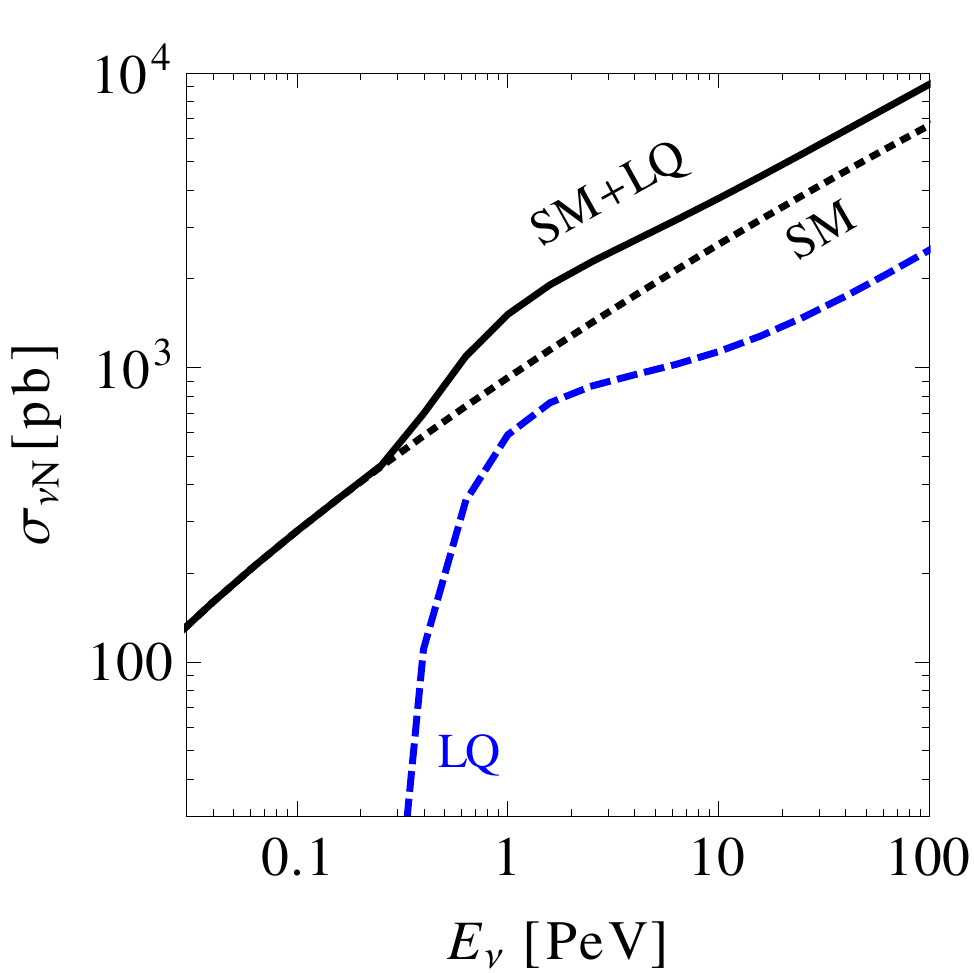}
\caption{(Left) $\nu_\tau$ neutrino-nucleon scattering cross-section from the SM and a 600 GeV leptoquark resonance with $y_{13}^L=1$. Both NC and CC interactions are included. $\nu_e$ and $\nu_\mu$ do not receive enhancement under our flavor assignment, and the leptoquark contribution to $\bar{\nu}_\tau$ is subdominant.}
\label{fig:xsec}
\end{figure}

At the IceCube detector, the incoming neutrinos with PeV energy would have enough center-of-mass (COM) energy $\sqrt{2 m _p E_\nu} \geq M_{LQ}$ to trigger a resonant $s$-channel leptoquark exchange. Due to the threshold neutrino energy for the leptoquark resonance, the neutrino-nucleon scattering cross-section deviates from that in the SM (as in Fig.~\ref{fig:xsec}), and can generally explain the emergence of the few PeV events with a coupling $y^L\sim 1$~\cite{Barger:2013pla}. In comparison, while an explanation using the SM only can also account for the observed number of PeV events due to its power-law spectrum which follows the shape of the incoming cosmic neutrinos, it would be difficult to develop features at a particular energy scale in the event energy spectrum. Admittedly the current data is statistics limited and the `discontinuous' spectral feature needs future measurements, the leptoquark resonance is a very interesting interpretation, and can be predictive when it is used to explain the CMS excess as well.

If the leptoquark interactions only involve  light quarks, the  differential resonance cross-sections are simplified compared to that in Ref.~\cite{Anchordoqui:2006wc},
\bea
\frac{d \sigma^{\text{NC/CC}}}{d y}&=&\frac{\pi}{2} {\cal R} \frac{D(M_{LQ}^2/s)}{s} ~,~\,\\
{\cal R}^{\text{CC}}&=& \frac{y_L^2(y_L^2+y_R^2)}{2y_L^2+y_R^2}~,~\,\\
{\cal R}^{\text{NC}}&=& \frac{y_L^4}{2y_L^2+y_R^2}~,~\,
\label{eq:LQdiffxsec}
\eea
where $y\equiv E_l/E_\nu$ is the daughter lepton energy fraction. $D$ is the averaged\footnote{With a 5:4 proton to neutron ratio in ice. The $d$ quark's proton distribution is taken to be same as  the $u$ quark inside a neutron.} nucleon's parton probability distribution at the resonance COM energy fraction $\hat{s}/s=M_{LQ}^2/s$. Depending on the leptoquark's decay channel, the analogue to the SM's charged-current (CC) final states $S\rightarrow lq$ deposit all the incoming energy into visible showers; in a neutral current (NC) $S\rightarrow \nu_l q'$ event the invisible neutrino hides away a fraction of the total energy. Therefore, the CC events are better candidates for the several highest-energy IceCube shower events, while the lower, sub-PeV ones could be a continuum due to NC events.

In terms of the couplings shown  in Table~\ref{tab:benchmarks}, we have $y_L^2=(y^L_{13})^2$ and $y_R^2=(y^R_{21})^2+(y^R_{23})^2$. It is clear that the CC process is less suppressed from sizable right-handed couplings compared to the NC process, as $S\rightarrow \nu_l q'$ is strictly left-handed. Fig.~\ref{fig:xsec} shows the SM and the leptoquark-enhanced neutrino cross-sections. Due to the appearance of the leptoquark resonance, the total cross-section exhibits a rise near the threshold neutrino energy $E_{\text{th}}=M_{LQ}^2/(2M_p)$. 

The cosmic neutrino flux is assumed to be isotropic,
\be 
\frac{d\phi}{dE_\nu}=3\phi_0 f_{i}\left(\frac{E_\nu}{100 \text{ TeV}}\right)^{-\gamma}~,~\,
\label{eq:sourcefunc}
\ee
where $f_{i}$ is the fraction of neutrinos of the $i$th flavor. The observed spectral parameters are given as $\gamma=2.6\pm0.15$ and $\phi_0=(2.3\pm0.4)\times 10^{-18}$ GeV$^{-1}$s$^{-1}$cm$^{-2}$sr$^{-1}$in Ref.~\cite{Aartsen:2015ivb}, which also reported good consistency with the conventional neutrino flavor ratio ($\frac{1}{3}:\frac{1}{3}:\frac{1}{3}$), and a best, however unphysical, fit at (0 : 0.2 : 0.8) with the current data. While muon track events only originate from $\nu_\mu$, the shower-events include contributions from the NC scattering from all favors as well as the CC scattering from both $\nu_e$ and $\nu_\tau$. For shower events, we only consider the `contained'-type of events in which the shower develops inside the Icecube's detector array volume. 
The contained shower event spectrum is
\bea 
\frac{dN}{dE}&=N_{\text{targets}} & \left(\int_{E}^{\infty}\frac{d\sigma^{NC}(E_\nu)}{dE}\frac{d\phi_\nu}{d E_\nu}dE_\nu \right. \nn \\
&&+\int_{E}^{\infty}\frac{d\sigma^{CC}(E_{\nu_\tau})}{dE}\frac{d\phi_{\nu_\tau}}{dE_{\nu_\tau}}dE_{\nu_\tau} \nn \\
&&+ \left.\sigma^{CC}(E)\frac{d\phi_{\nu_e}}{d E} \right) ~.~\,
\label{eq:showerEvt}
\eea
The leptoquark component in the NC scattering cross-section can be easily derived from Eq.~(\ref{eq:LQdiffxsec}) as ${d\sigma/dE}=1/E_{\nu}\cdot d\sigma(E_\nu)/dy$, where $E$ is the visible energy deposit. The $\nu_e$ CC scattering deposits 100\% of the neutrino energy into showers, while for $\nu_\tau$ CC scattering, a loss in the energy deposition occurs due to the secondary neutrinos that emerge from $\tau$ decays. % see Appendix~\ref{app:nutauCC} for details.
The CC showers from $\nu_e$ and $\nu_\tau$ have unequal hadronic contents and consequently generate different amount of light yields. A full shower acceptance simulation is beyond the scope of this paper, and we adopt the approximation given in Ref.~\cite{Kistler:2013my} that 100\% of  $\nu_e$ CC shower is measured, while for $\nu_\tau$ it is approximated as $E\approx 0.5E_{\nu_\tau}$, or $d\sigma/dE=\sigma^{CC}(E_{\nu_\tau})\cdot \delta(E-E_{\nu_\tau}/2)$. Similarly, for the mainly hadronic NC showers, we use $d\sigma/dE=\sigma^{NC}(E_{\nu})\cdot\delta(E-0.2 E_{\nu})$ for all flavors. 

Just like the SM, both  CC and NC processes contribute. However,{\it ~unlike} the SM, 
the leptoquark resonance only occurs involving a neutrino and a quark, while the CP-conjugate process, the anti-neutrino resonance scattering is highly suppressed due to the anti-quarks being only sea quarks inside the nucleons. The parametrizations given in Table~\ref{tab:benchmarks} only needs to include the leptoquark enhancement in the $\nu_\tau$ scattering.

\begin{figure}[h]
\includegraphics[width=8cm,height=5cm]{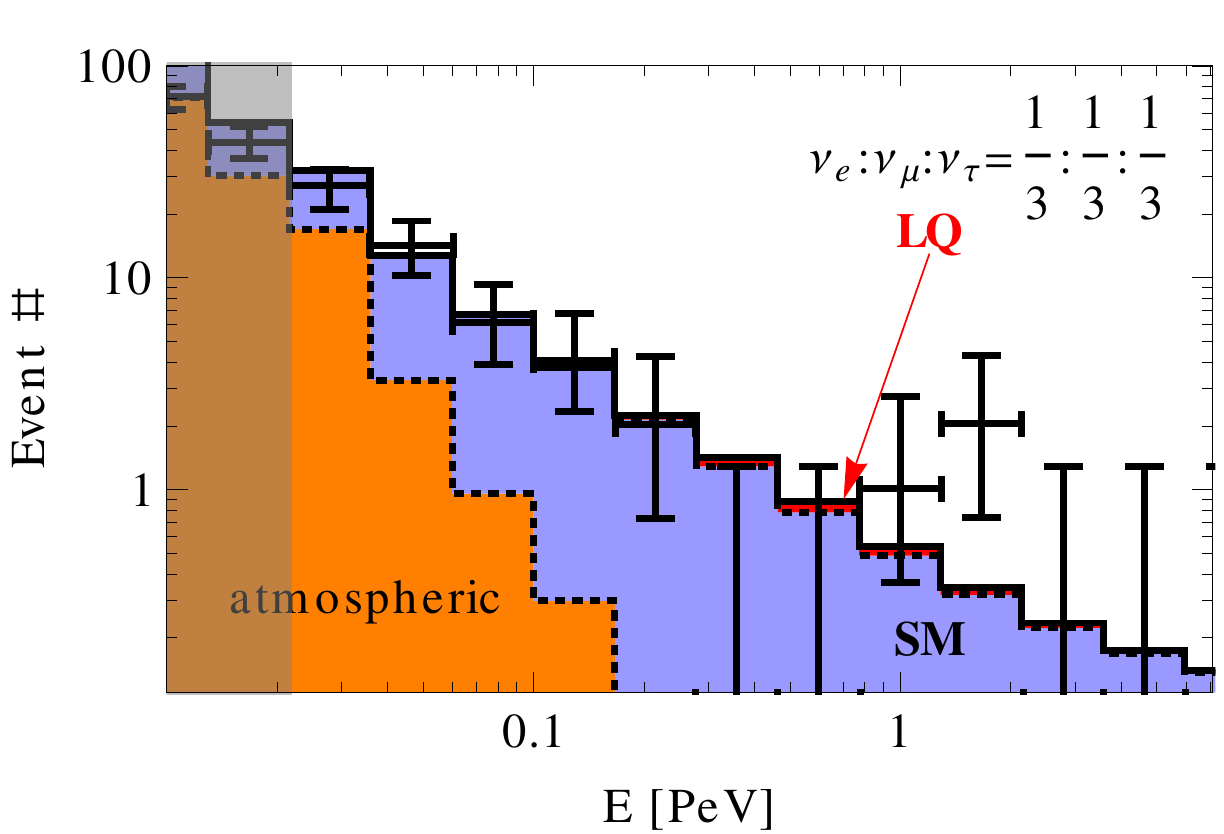}
\includegraphics[width=8cm,height=5cm]{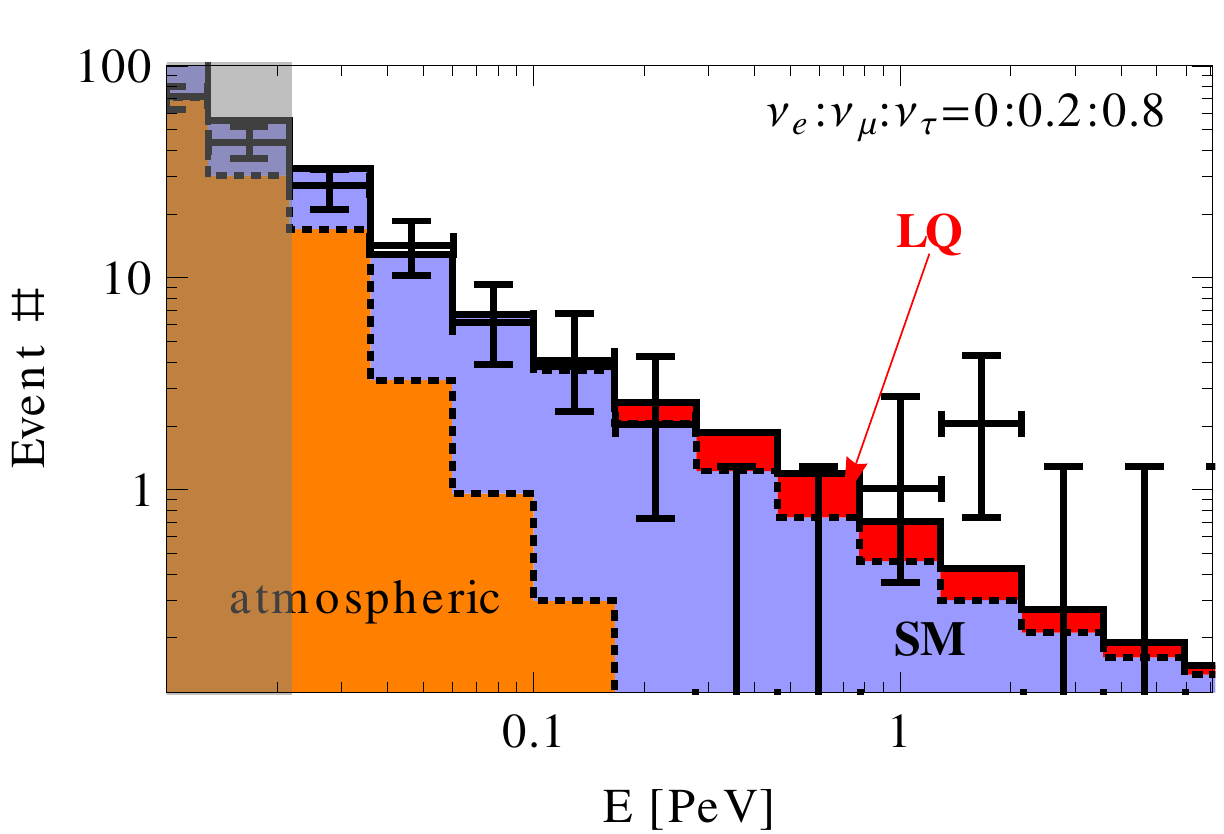}
\caption{The shower event rates in the southern hemisphere. The leptoquark contribution assumes $M_{LQ}=600$ GeV and $y_{13}^L=1.4$. The first panel assumes a ($\frac{1}{3}:\frac{1}{3}:\frac{1}{3}$) flavor ratio while the second panel assumes the best-fit ($0:0.2:0.8$) ratio, which makes the LQ signal more pronounced with less SM $\nu_e$ CC showers.}
\label{fig:SHemiShowers}
\end{figure}

Fig.~\ref{fig:SHemiShowers} shows the leptoquark's enhancement to PeV shower events in the southern hemisphere. We show both the conventional 
($\frac{1}{3}:\frac{1}{3}:\frac{1}{3}$) and the best-fit (0:0.2:0.8) flavor ratio. In the latter case the leptoquark signal becomes enhanced due to its mostly third flavor coupling. Since the leptoquark resonance only occurs in the PeV events, the energy dependence in the effective volume is dropped, which would be more significant around or below tens of TeV. A constant (saturated) detector acceptance is assumed and the SM event rate is calibrated to the sub-PeV data. The red shaded region assumes a coupling size $y_{13}^L=1.4$ for the first benchmark scenario in Table~\ref{tab:benchmarks}. The second benchmark results in a very similar signal and is not shown. The majority of the leptoquark signals occur via $\nu_\tau$ CC scattering, which would indicate itself through a rise in the fraction of $\tau$s in the PeV events.

Interestingly, the significance of the leptoquark contribution can be sensitive to the neutrino flavor composition under the fractional energy deposition scheme in Eq.~\ref{eq:showerEvt}, where $\nu_e$ CC presumably  deposits 100\% neutrino energy and becomes the most efficient contributor to the PeV event rate because it picks up cosmic flux at a lower energy, which falls steeply as an $E_\nu^{-2.6}$ power-law. As a result, the SM shower rate becomes dependent on the $\nu_e$ composition. As shown in the lower panel in Fig.~\ref{fig:SHemiShowers}, the ($0:0.2:0.8$) flavor ratio, which would require beyond SM sources, leads to a smaller SM rate relative to an enhanced (by a higher $\nu_\tau$ composition) leptoquark rate, making the leptoquark  more visible. The lower $\nu_e$ CC contribution also causes a difference in the 
effective exposure when calibrating to the sub-PeV SM events, which is 650 Megaton of water-equivalent times year for 
the ($\frac{1}{3}:\frac{1}{3}:\frac{1}{3}$) flavor ratio and \sci{1.3}{3} Mton$\cdot$year for ($0:0.2:0.8$), at the central  values of parameters used in Eq.~\ref{eq:sourcefunc}. 

{\bf As a caveat}, this effect from neutrino composition might be over-estimated because $\nu_\tau$ CC and the NC events in principle can deposit more than their average fractions of energy in $d\sigma(E_\nu)/dE$. A proper detector simulation would then be required, especially in the case of $\nu_e$-depleted sources, to make a more precise prediction. Also, while a leptoquark enhancement to $\nu_\tau$ scattering may make the source appear to have more $\nu_\tau$ in a SM-only calculation, the relatively small leptoquark event rate (as seen in the upper panel of Fig.~\ref{fig:SHemiShowers}) with a ($\frac{1}{3}:\frac{1}{3}:\frac{1}{3}$) flavor is insufficient to boost the cosmic source into a ($0:0.2:0.8$) appearance. So we utilize the latter only to demonstrate the leptoquark's sensitivity on the flavor composition of the neutrino source flux. %As IceCube's event statistics increase, this best-fit flavor ratio may change, as is recently reported in Ref.~\cite{bib:IPA2015OlgaBotner}. %https://events.icecube.wisc.edu/conferenceTimeTable.py?confId=68#20150504

In Fig.~\ref{fig:SHemiShowers}, we illustrated with $|y^L_{13}|^2=2$. The leptoquark event rate at other coupling values can be simply scaled by $|y|^2$. The enhancement  raises the event rate over a wide range of energy. More events are expected  just above the threshold, as can be seen in Fig.~\ref{fig:xsec}. In the high energy limit, the leptoquark contribution approaches to a constant enhancement and the total event spectrum would return to a power law which follows the shape of the incoming cosmic flux.

\section{discussions}
\label{sect:conclusion}

It is clear that  the presence of a leptoquark 
helps to fit the IceCube data even though the inclusion of the leptoquark alone may not fully explain the shape of the PeV shower spectrum with only a few events at present, where the SM still plays a major role.  If excessive PeV shower events keep occuring in the future data, a change in the power-index of the event energy spectrum at the PeV energy scale can be interpreted as a leptoquark signal. %Since the CC process dominates leptoquark's contribution to the shower events, the introduction of large right-handed couplings, as the CMS excess requires, results in a slightly higher event rate compared to purely left-handed case. The NC process, however, becomes very subdominant.
IceCube's discovery analysis uses a very  conservative selection criteria to extract a very low backgrounds sample, given that the flux was unknown. With a better understanding of the signal, re-optimized selection criteria can improve event statistics significantly, for example, by relaxing the containment criteria. Further, the interest in gigaton size detectors has substantially increased, with several projects targeting the high energy region, e.g.,  IceCube-Gen2~\cite{bib:ICgen2}, KM3Net in the Mediterranean~\cite{bib:KM3Net}, and Gigaton Volume Detector (GVD)~\cite{bib:GVD} in lake Baikal. By the end of this decade, IceCube alone will  triple  the sample size (by the most conservative estimates) in the PeV region due to the increased exposure. In the early next decade, using upgrades like Gen2 alone or in combination with GVD and KM3Net, one could easily obtain a ten fold increase in the annual event rate and a precise measurement of the PeV spectrum would then allow us to distinguish between conventional  and leptoquark scenarios.

If the PeV events were enhanced by a leptoquark around 600-650 GeV, the CMS data would require this leptoquark left-handedly couple to the 3rd generation leptons and the first generation quarks, also with sizeable right-handed couplings, as represented by the benchmark points. This would predict more $\nu_\tau$ events at PeV. However, a fairly large coupling $y_L\ge 1$ is required for the leptoquark contribution to $\nu N$ scattering to be comparable to that from the SM. The upcoming LHC data in the $eejj$ and $e\nu jj$ channels can easily verify or exclude leptoquark couplings of such a magnitude, as well as the flavor structures shown in Table~\ref{tab:benchmarks}.

\section{Acknowledgement}

We thank Jason Evans and Natsumi Nagata for communications on leptoquarks's correction to the meson decay and $u$-quark mass. The work of B.D. is supported by DOE Grant DE-FG02-13ER42020. Y.G. thanks the Mitchell Institute for Fundamental Physics and Astronomy for support. The work of T.L. is supported by 
the Natural Science Foundation of China under grant numbers 11135003, 11275246, and 11475238,  
and by the National Basic Research Program of China (973 Program) under grant number 2010CB833000.
The work of C.R. is supported by the Basic Science Research Program through the National Research Foundation of Korea funded by the Ministry of Science, NRF-2013R1A1A1007068. 
The work of L.E.S. supported by NSF Grant No. PHY-1522717.

\end{document}